\pdfoutput=1

\documentclass[10pt,conference]{IEEEtran}

\usepackage[T1]{fontenc}
\usepackage[latin9]{inputenc}
\usepackage{dirtytalk}
\usepackage[detect-all]{siunitx}
\usepackage[inline]{enumitem}
\usepackage[hyphens]{url}
\usepackage{booktabs}
\usepackage[capitalize,noabbrev]{cleveref}
\usepackage{support-caption}
\usepackage{subcaption}
\usepackage{graphicx}
\graphicspath{{images/}}

\newcommand{\cor}{code review}
\newcommand{\cov}{code velocity}
\newcommand{\gh}{{G}it{H}ub}
\newcommand{\fbsd}{{F}ree{BSD}}
\newcommand{\llvm}{\textsc{LLVM}}
\newcommand{\mcr}{{M}odern {C}ode {R}eview}
\newcommand{\mo}{{M}ozilla}
\newcommand{\oss}{\textsc{OSS}}
\newcommand{\sloc}{\textsc{SLOC}}
\newcommand{\tta}{time-to-accept}
\newcommand{\ttfr}{time-to-first-response}
\newcommand{\ttm}{time-to-merge}
\newcommand{\ger}{{G}errit}
\newcommand{\phab}{{P}habricator}

\newcommand{\InitialPhabricatorCount}{\num{283235}}
\newcommand{\FilteredPhabricatorReviews}{\num{280456}}
\newcommand{\MedianBlenderCommits}{\num{8589}}
\newcommand{\MedianLLVMCommits}{\num{ 29555}}
\newcommand{\MedianBlenderDistinctAuthors}{\num{89}}
\newcommand{\MedianLLVMDistinctAuthors}{\num{592}}
\newcommand{\AverageProjectAge}{seven}
\newcommand{\AnnualSLOCIncrease}{\num{3}--\num{17}\%}
\newcommand{\anonymize}[2]{#2}

\makeatletter

\def\lst@makecaption{
  \def\@captype{table}
  \@makecaption
}

\makeatother

\newenvironment{boxed_new}
    {\begin{center}
    \begin{tabular}{|p{0.45\textwidth}|}
    \hline\\
    }
    {
    \\\\\hline
    \end{tabular}
    \end{center}
    }

\makeatletter

\begin{document}

\title{Are We Speeding Up or Slowing Down?\\On Temporal Aspects of Code Velocity}

\author{
    \anonymize{
        \IEEEauthorblockN{
            Anonymous Author(s)
            \bigskip
            \bigskip
            \bigskip
        }
    }{
    \IEEEauthorblockN{Gunnar Kudrjavets}
    \IEEEauthorblockA{
        \textit{University of Groningen}\\
                9712 CP Groningen, Netherlands \\
                g.kudrjavets@rug.nl}

    \and

    \IEEEauthorblockN{Nachiappan Nagappan}
    \IEEEauthorblockA{
        \textit{Meta Platforms, Inc.} \\
                Menlo Park, CA 94025, USA \\
                nnachi@meta.com}

    \and

    \IEEEauthorblockN{Ayushi Rastogi}
    \IEEEauthorblockA{
        \textit{University of Groningen}\\
                9712 CP Groningen, Netherlands \\
                a.rastogi@rug.nl}
}}

\maketitle

\IEEEtriggeratref{19}

\begin{abstract}
This paper investigates how the duration of various \cor\ periods changes over a projects' lifetime.
We study four open-source software (\oss) projects: Blender, \fbsd, \llvm, and \mo.
We mine and analyze the characteristics of \InitialPhabricatorCount\ \cor s that cover, on average, \AverageProjectAge\ years' worth of development.
Our main conclusion is that neither the passage of time or the project's size impact \cov.
We find that
\begin{enumerate*}[label=(\alph*),before=\unskip{ }, itemjoin={{, }}, itemjoin*={{, and }}]
    \item the duration of various \cor\ periods (\ttfr, \tta, and \ttm) for \fbsd, \llvm, and \mo\ either becomes shorter or stays the same; no directional trend is present for Blender
    \item an increase in the size of the code bases (annually \AnnualSLOCIncrease) does not accompany a decrease in \cov
    \item for \fbsd, \llvm, and \mo, the \num{30}-day moving median stays in a fixed range for \ttm.
\end{enumerate*}
These findings do not change with variabilities in code churn metrics, such as the number of commits or distinct authors of code changes.
\end{abstract}


\section{Introduction}

One critical goal in the software industry is to develop, review, integrate, and deploy code changes \emph{fast}.
The software industry focuses on \emph{increasing the \cov}~\cite{chandra_2022,riggs_move_2022,chen_2022} using different tools or process enhancements.
Various adaptations of Continuous Integration (\textsc{CI}) and Continuous Deployment (\textsc{CD})~\cite{fowler_continuous_2006} have become default practices for most of the current projects in both industry and \oss\ communities.
Similarly, the lightweight \mcr~\cite{bird_2015,sadowski_modern_2018}  is now a de facto standard process to conduct \cor s.
For existing non-agile projects,
the initial switch from methodologies such as the waterfall model to \textsc{CI}/\textsc{CD} or mailing lists to contemporary code collaboration tools can have an immediate and noticeable impact on increasing \emph{code velocity}~\cite{jha_2016,kaur_2021}.
The longevity of these improvements to \cov\ has yet to be thoroughly investigated.

We focus on studying the direction of a change in the duration of various \cor\ periods as the software projects evolve.
We intend to determine how the speed of \cor s changes over time.
We mine \cor\ data from four different \oss\ projects:
Blender~\cite{blender_phabricator},
\fbsd~\cite{freebsd_phabricator},
\llvm~\cite{llmv_phabricator},
and \mo~\cite{mozilla_phabricator}.
We analyze \InitialPhabricatorCount\ \cor s and evaluate how different \cor\ periods (\ttfr, \tta, and \ttm) trend over time.
Our study suggests that the speed of \cor s remains the same as the projects evolve.

\section{Background and motivation}
\label{sec:background}

\begin{table*}[htbp]
    \caption{The \textit{p} values from the modified Mann-Kendall (\textsc{MK}) test and magnitude of change for Sen's slope (Theil-Sen estimator).
    We present the 30-day moving median and all-time median for different \cor\ periods.
    Statistically significant Mann-Kendall \textit{p} value indicates the presence of either an upward or downward monotonic trend.
    Sen's slope (presented at a 95\% confidence level) identifies the magnitude of the trend per unit time step.
    The time step is \num{30} days.
    The unit for \cor\ periods is hours.}
    \label{tab:MK_and_Sens}
    \centering

    \begin{tabular}{lrrrrrrrrrrrr}
        \toprule
        \multicolumn{1}{l}{} &
        \multicolumn{4}{c}{\textbf{Time-to-first-response}} &
        \multicolumn{4}{c}{\textbf{Time-to-accept}} &
        \multicolumn{4}{c}{\textbf{Time-to-merge}}
        \\
        \cmidrule(lr){2-5}
        \cmidrule(lr){6-9}
        \cmidrule(lr){10-13}
        &
        \multicolumn{2}{c}{30-day median} &
        \multicolumn{2}{c}{All-time median} &
        \multicolumn{2}{c}{30-day median} &
        \multicolumn{2}{c}{All-time median} &
        \multicolumn{2}{c}{30-day median} &
        \multicolumn{2}{c}{All-time median}
        \\
        \cmidrule(lr){2-3}
        \cmidrule(lr){4-5}
        \cmidrule(lr){6-7}
        \cmidrule(lr){8-9}
        \cmidrule(lr){10-11}
        \cmidrule(lr){12-13}
        \textbf{Name}&
        \textsc{MK} & Sen's &
        \textsc{MK} & Sen's &
        \textsc{MK} & Sen's &
        \textsc{MK} & Sen's &
        \textsc{MK} & Sen's &
        \textsc{MK} & Sen's
        \\
        \cmidrule(lr){1-1}
        \cmidrule(lr){2-2}
        \cmidrule(lr){3-3}
        \cmidrule(lr){4-4}
        \cmidrule(lr){5-5}
        \cmidrule(lr){6-6}
        \cmidrule(lr){7-7}
        \cmidrule(lr){8-8}
        \cmidrule(lr){9-9}
        \cmidrule(lr){10-10}
        \cmidrule(lr){11-11}
        \cmidrule(lr){12-12}
        \cmidrule(lr){13-13}
        Blender &
        $.231$  &                &
        $.229$  &                &
        $.715$  &                &
        $.846$  &                &
        $.609$  &                &
        $.666$  &                \\
        FreeBSD &
        $<.001$ & $-\num{0.003}$ &
        $<.001$ & $-\num{0.002}$ &
        $<.001$ & $-\num{0.011}$ &
        $<.01$  & $-\num{0.003}$ &
        $.193$  &                &
        $<.05$  & $\num{0.006}$  \\
        \textsc{LLVM} &
        $<.001$ & $\num{0.000}$  &
        $<.001$ & $\num{0.000}$  &
        $<.001$ & $-\num{0.005}$ &
        $<.001$ & $-\num{0.002}$ &
        $<.001$ & $-\num{0.009}$ &
        $<.001$ & $-\num{0.002}$ \\
        Mozilla &
        $<.001$ & $-\num{0.001}$ &
        $<.001$ & $-\num{0.001}$ &
        $<.01$  & $\num{0.000}$  &
        $<.001$ & $\num{0.000}$  &
        $<.01$  & $-\num{0.001}$ &
        $.060$  &               \\
        \bottomrule
    \end{tabular}
\end{table*}

Different interpretations and scopes for \cov\ exist.
A general definition is \say{the time between making a code change and shipping the change to customers}~\cite{code_vel_definition_ast}.
In this paper, we focus on a more quantifiable metric related to the duration of \cor s.
We use \emph{\ttm} as a proxy metric to quantify how fast code changes propagate.
The \ttm\ covers a period from when an engineer publishes a set of code changes that are ready for \cor\ till these changes are merged to the target branch~\cite{izquierdo-cortazar_2017}.

\emph{There are two prevailing and contradictory theories within industry about the direction of \cov\ over time}.
The first hypothesis states that due to an increase in the size and complexity of a system (Lehman's second law \say{Increasing Complexity}~\cite{lehman_1980}), e.g., increase in the complexity of communication due to a bigger team, \cov\ decreases.
A second hypothesis argues that \cov\ increases over time.
The increase in \cov\ is because engineers become more familiar with the code base, interpersonal communication becomes more efficient, and the tooling infrastructure improves.

Another of Lehman's laws of software evolution is \say{Continuing Growth}~\cite{lehman_1980}.
Lehman's sixth law states that the \say{[f]unctional content of a program must be continually increased to maintain user satisfaction over its lifetime}~\cite{lehman_1991_6th_law}.
As a result of additional functionality, it is reasonable to assume that the size of the code base increases.
The size of the code base is typically measured in source lines of code (\sloc).
Based on our industry experience, we also observe that in conjunction with the new demands on a project, the size of the development teams tends to increase rather than decrease.
According to Brooks' law, \say{[a]dding manpower to a late software project makes it late}~\cite{brooks_1995}.
While the software project does not necessarily have to be late while developing new features, it is unknown how increased code base and team sizes impact \cov.
To investigate this subject further, we formulate the following research question:

\begin{boxed_new}
    \textbf{RQ:} How does a project's \cov\ trend over time? Does the \cov\ increase, decrease, or stay neutral?
\end{boxed_new}

\section{Methodology}

\subsection{Choice of data}

\Cref{sec:background} states that our primary \cov\ metric of interest is \ttm~\cite{izquierdo-cortazar_2017}.
We also know that research from both Google~\cite{google_reviews_2022} and
Microsoft~\cite{bird_2015} finds that \ttfr\ and \tta\ are additional \cor\ metrics important for developers.

Existing \cor\ datasets focus either on \gh~\cite{gousios_ghtorrent_2013} or \ger~\cite{mukadam_2013,paixao_2018,yang_2016}.
By default, \ger\ immediately merges changes once they are accepted.
That behavior means that treating \tta\ and \ttm\ as separate events is neither valuable nor valid.
\gh\ added the ability to approve changes only in 2016~\cite{github_code_review_2016}.
We randomly chose \num{100} \gh\ projects and inspected pull requests in those projects.
The adoption rate and consistent use of that feature still need to be higher to yield valuable data.

One code collaboration tool that exposes a data model that formally tracks various \cor\ periods is \phab~\cite{phacility}.
Based on public information about existing \phab\ projects~\cite{phabricator_usage},
we mine data for four major \oss\ projects with a multi-year development history.
Those projects are Blender, \fbsd, \llvm, and \mo.
Our initial dataset contains \InitialPhabricatorCount\ \cor s.
We removed the \cor s where the \mcr\ did not happen.
Examples of nonconforming \cor s are the ones
that do not contain any lines of code,
where the author has accepted their changes,
or \cor s that were committed without any acceptance.
After filtering and applying consistency checks, the final dataset contains \FilteredPhabricatorReviews\ \cor s.

\begin{figure*}[htbp]
  \caption{The trend of 30-day rolling medians of \ttm\ for different projects.}
  \label{fig:30-day-trend}
  \centering
  \begin{subfigure}{0.45\textwidth}
    \includegraphics[width=\textwidth]{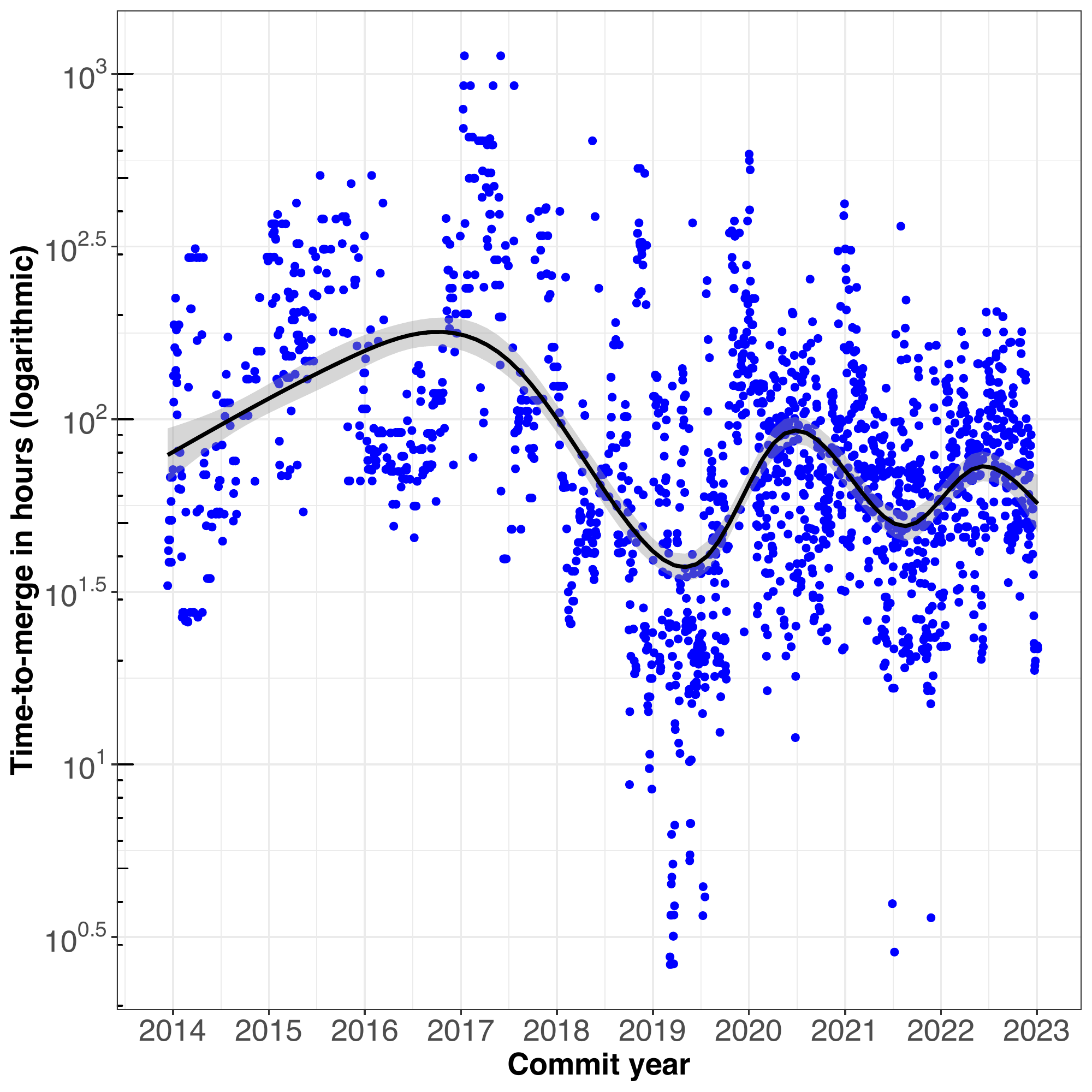}
    \caption{Blender.}
  \end{subfigure}
  \hfill
  \begin{subfigure}{0.45\textwidth}
    \includegraphics[width=\textwidth]{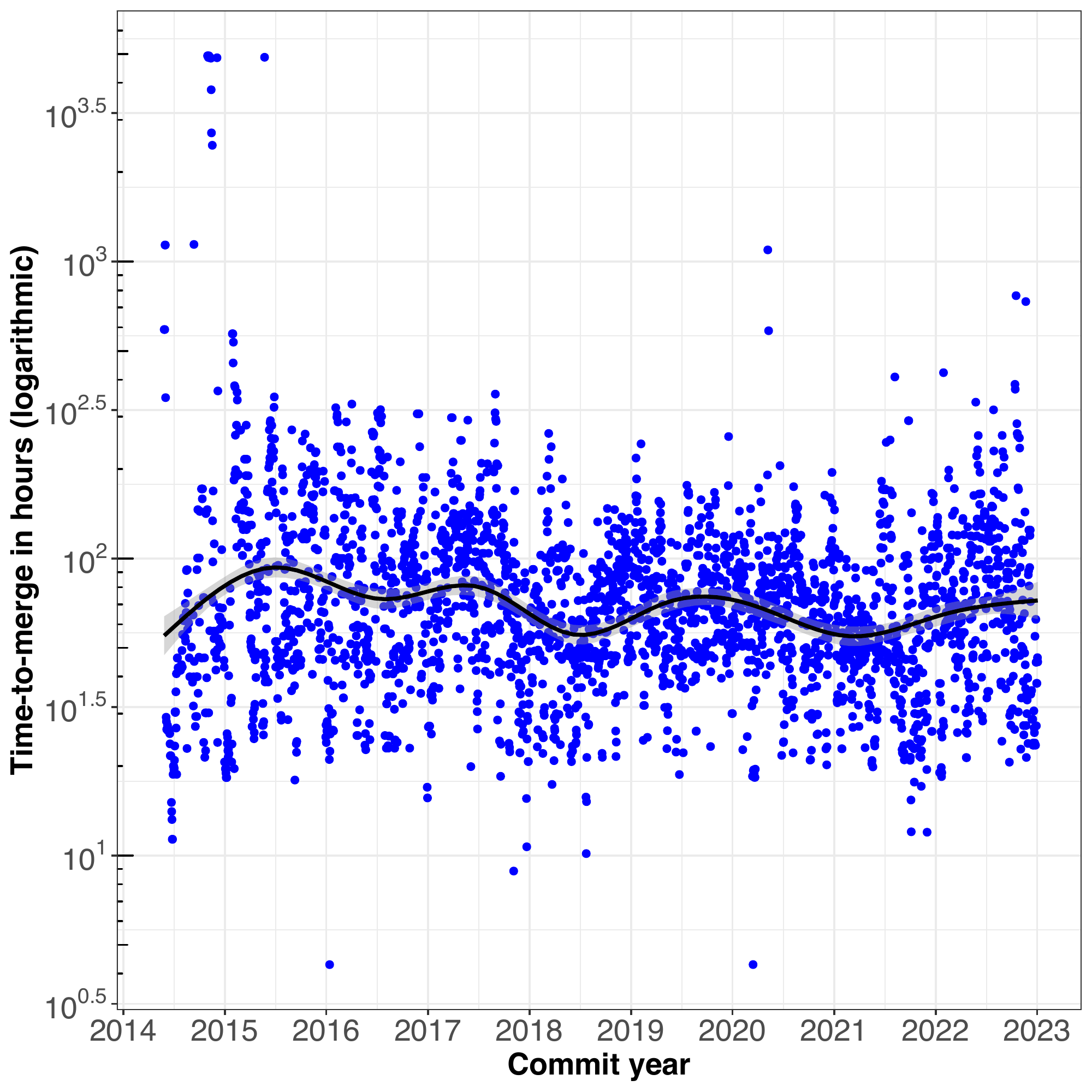}
    \caption{FreeBSD.}
  \end{subfigure}

  \begin{subfigure}{0.45\textwidth}
    \includegraphics[width=\textwidth]{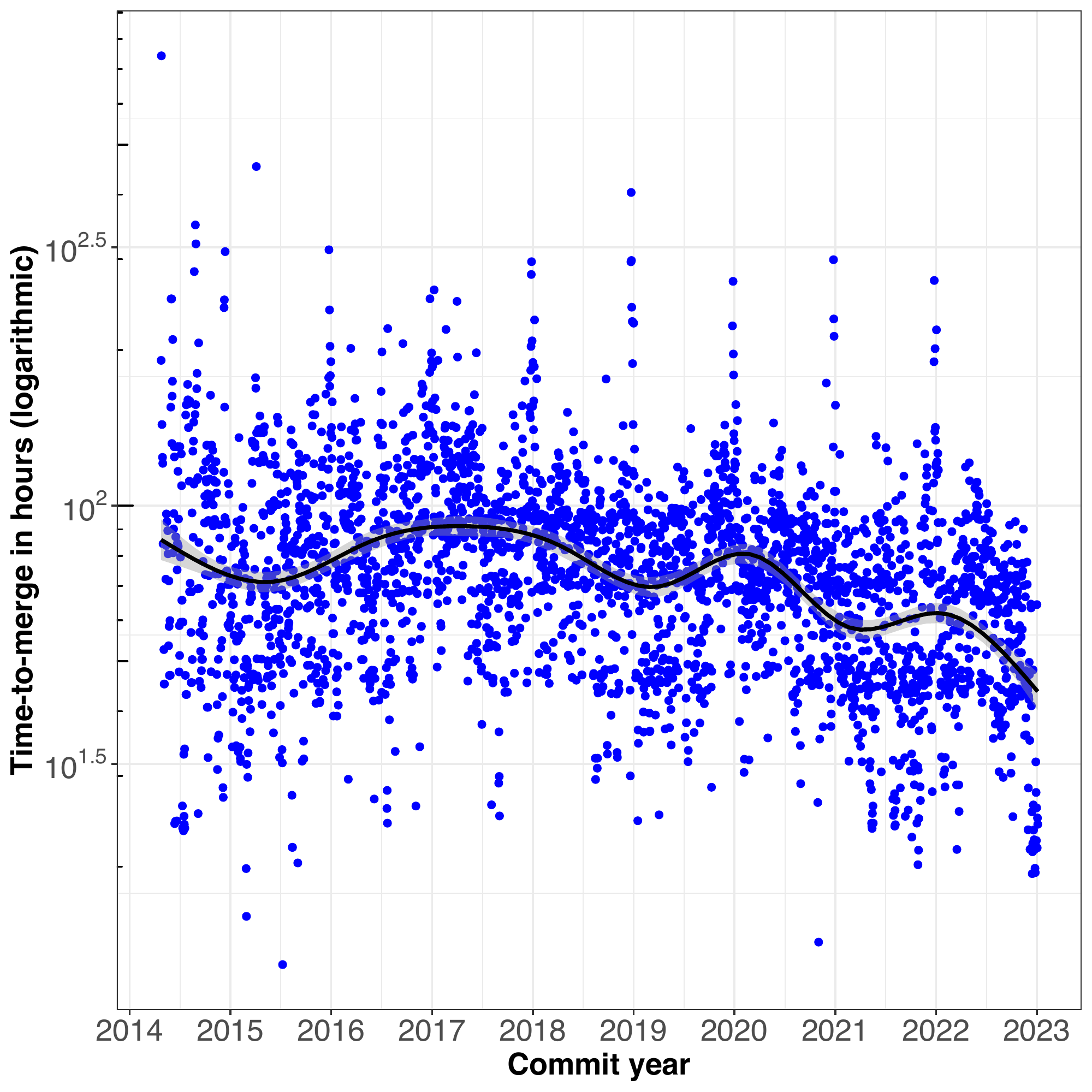}
    \caption{\textsc{LLVM}.}
  \end{subfigure}
  \hfill
  \begin{subfigure}{0.45\textwidth}
    \includegraphics[width=\textwidth]{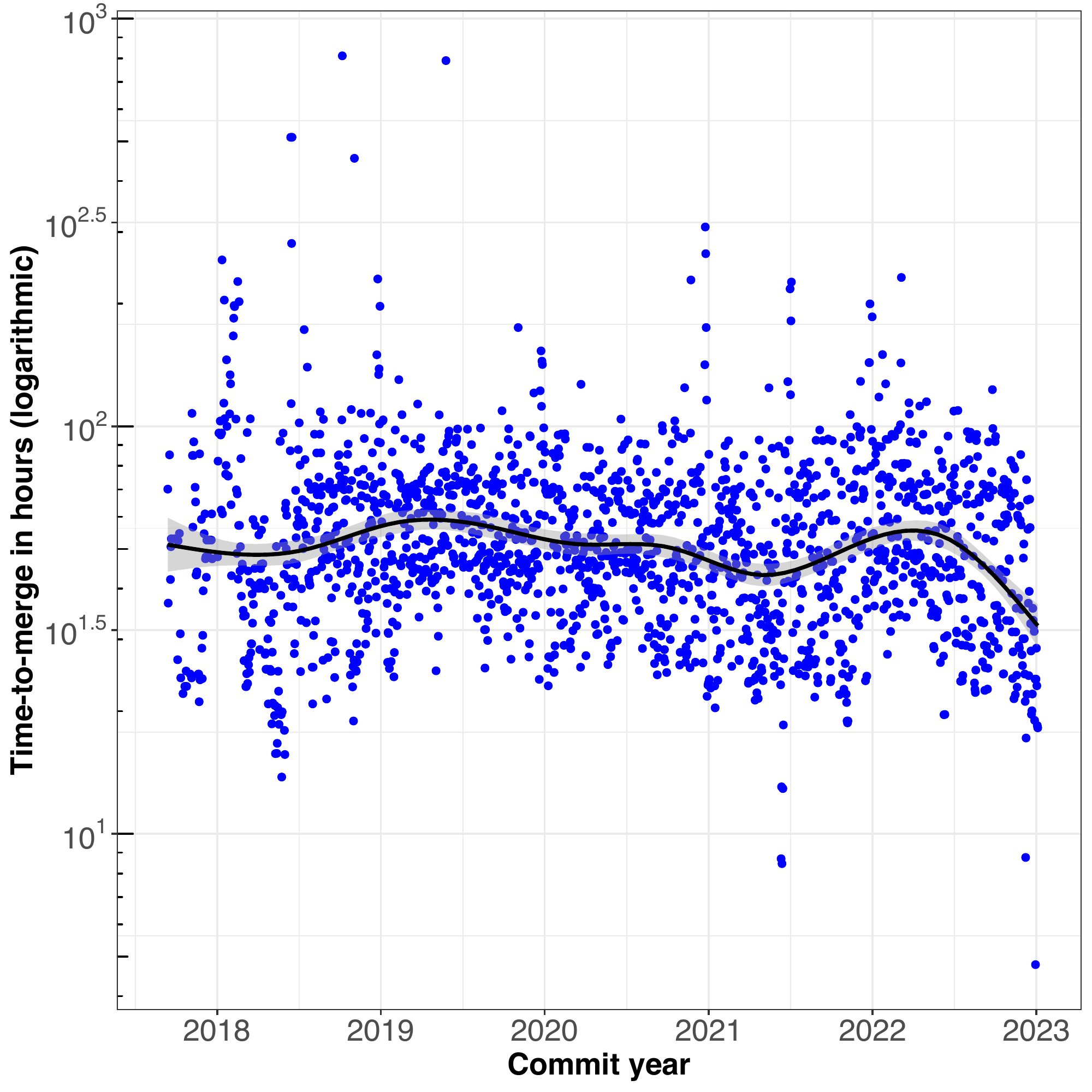}
    \caption{Mozilla.}
  \end{subfigure}
\end{figure*}

\subsection{Statistical analysis}

As a first step, we investigate if there is a trend in \cov.
We use the Mann-Kendall test~\cite{mann_1945,kendall_1976} to determine if there is a monotonic upward or downward trend.
The monotonic trend~\cite{hirsch_1981} means that \say{the variable consistently increases (decreases) through time, but the trend may or may not be linear}~\cite{mann-kendall_trend}.
The null hypothesis ($H_0$) for the Mann-Kendall test is that there is no monotonic trend.
The alternative hypothesis ($H_1$) is the presence of a monotonic trend.
Secondly, if a statistically significant Mann-Kendall correlation is present, then we calculate the Sen's slope (Theil-Sen estimator)~\cite{sen_1968,gilbert_1987} to evaluate the magnitude of the trend.
Sen's slope indicates the rate of change per unit time step.

How to handle autocorrelation is a challenge for time series analysis.
One popular approach is to aggregate the time series to use coarser time granularities such as monthly or yearly samples~\cite{collaudcoen_2020}.
Another mitigation is to use a modified Mann-Kendall test to adjust for autocorrelation~\cite{hamed_1998,yue_2002}.
We use both techniques to reduce the chance of falsely concluding that a trend is present when it is not.

We conduct these calculations for three \cor\ periods: \ttfr, \tta, and \ttm.
We look at the 30-day rolling (moving) and all-time median.
The all-time median is the median of the specific metrics up to a given time.
Both metrics help to evaluate the trend of a specific variable~\cite{hyndman_2011}.
We chose the median values as opposed to the mean values because the median is more resistant to outliers.
We use an $\alpha$ level of $.05$ for our statistical tests.

\section{Results}

The results from the statistical analysis are displayed in \Cref{tab:MK_and_Sens}.
The main observations from that analysis are the following:
\begin{enumerate*}[label=(\alph*),before=\unskip{ }, itemjoin={{, }}, itemjoin*={{, and }}]
    \item
    except for Blender, there is a statistically significant monotonic trend for all the projects
    \item
    while a statistically significant trend is present, based on its numeric values and visual representation (see~\Cref{fig:30-day-trend}), it is minimal
    \item
    all the 30-day median slope values are negative or zero, which suggests a \emph{minor increase in \cov}.
\end{enumerate*}

Using a \num{30}-day rolling mean or median is standard practice.
Depending on the context, \num{90}-day (quarterly) technical indicators are also used to check for trends.
We also calculate the Mann-Kendall and Sen's slope values as an additional data point using the \num{90}-day rolling median.
The conclusions do not change as a result.

Complimentary to the analysis above, the visualization of the trend for \num{30}-day rolling median \ttm\ is displayed in~\Cref{fig:30-day-trend}.
Based on visual observation, we note that for \fbsd, \llvm, and \mo, most median values stay in a relatively fixed range.

For Blender, there is a noticeable cluster of results that indicate increased \cov\ between 2017 and 2019.
To further investigate this result, we inspected \num{50} random Blender \cor s between 2017 and 2019.
We cannot find conclusive evidence that explains the drop in \num{30}-day rolling median values.
Based on the public information (\say{Blender now has a much larger team of people working on core development}), we speculate that a sudden increase in the number of engineers may have caused a temporary increase in \cov~\cite{roosendaal_blender_2019}.
The trend may have normalized in 2020 because of \say{unprecedented number of 108 new contributors}~\cite{siddi_blender_2021}.

Various data points and events can influence a project's \cov.
Only a few of these potential variables are formally tracked.
Decisions about feature development, project management, organizational challenges, or changing business priorities are not always documented and available to the public.
We have only limited insight into all the confounding variables.
To understand the potential impact of metrics associated with project development, we mine the ones we can access.
Our findings are displayed in~\Cref{tab:project-growth-rate}.
The only common indicator between the projects is the continuous increase in the source lines of code.
We focus on \emph{objective characteristics} that we can mine from the source control management system and defect database.
We look at the annual change in
total source lines of code (calculated using scc~\cite{boyter_scc}),
number of commits,
number of distinct authors, and
number of distinct committers.
We use each project's default branch
from the original Git repository or an available \gh\ read-only mirror.

\begin{table}[htbp]
    \caption{Median annual increase or decrease percentage in various code churn metrics per project. The scope is a \cor\ period covered in~\Cref{fig:30-day-trend}. We separate the roles of an author and a committer because they can differ.}
    \label{tab:project-growth-rate}
    \centering

    \begin{tabular}{llrrrr}
        \toprule
        \textbf{Project} & \textbf{Period} & \textbf{\sloc} & \textbf{Commit} & \textbf{Distinct} & \textbf{Distinct}  \\
        &  &  & \textbf{count} & \textbf{author} & \textbf{committer}  \\
        \midrule
Blender & 2014--2022 & $\num{6.2}\%$ &$\num{-2.7}\%$ &$\num{13.5}\%$ &$\num{2.3}\%$ \\
FreeBSD & 2014--2022 & $\num{3.5}\%$ &$\num{-1.2}\%$ &$\num{0.5}\%$ &$\num{-4.3}\%$ \\
\textsc{LLVM} & 2014--2022 & $\num{17.4}\%$ &$\num{6.8}\%$ &$\num{18.5}\%$ &$\num{15.6}\%$ \\
Mozilla & 2017--2022 & $\num{11.1}\%$ &$\num{-4.8}\%$ &$\num{-6.9}\%$ &$\num{-13.8}\%$ \\
       \bottomrule
    \end{tabular}
\end{table}

We investigate the contents of defect-tracking databases associated with the projects we analyze.
We find the metrics related to defects to be an unreliable indicator of the project's workload.
We find no formal rules related to defect counts and their implication for the development process.
In addition, we do not observe that developers use the defects assigned to them as a primary list of work items.

\section{Discussion}

Our main finding is that \emph{there is no significant change in the trend for various \cor\ periods}.
The lack of trend applies to all the projects regardless of the increase or decrease in commit count, number of distinct authors, and number of distinct committers.
The median annual change in code base size for all the projects is an increase of \AnnualSLOCIncrease.
At the same time, the \num{30}-day median and all-time median for different \cor\ periods change only in a second or third decimal place (see~\Cref{tab:MK_and_Sens}).
If anything, there is a minor decrease in the duration for \ttfr, \tta, and \ttm.

The main finding is surprising.
The conventional wisdom in software engineering is that communication and processes slow down with an increase in the project's size and age.
There could be multiple explanations for the results that we see.
One possibility is that while the code complexity and team size continue to increase, engineers get more familiar and efficient while working in the project's code base.
With time the developers also build better interpersonal relationships that improve communication efficiency.
It is reasonable to assume that development infrastructure
also improves.
Those factors can counteract the time spent on tasks such as complex debugging issues or the comprehension speed for the code sent for review.

Another possibility is that \emph{regardless of the project size, the \cov\ stays in a specific range due to factors such as human nature}.
Engineers will get to reviewing the code when they get to it.
Developers will spend a fixed amount of time on \cor s regardless of their familiarity with the code or other parallel priorities.
The \say{flatness} or lack of a trend is visible in~\Cref{fig:30-day-trend} for \fbsd, \llvm, and \mo.
The Blender project has more variability in \ttm.
The variability can be explained by Blender being the smallest of all the projects we investigate.
For example, the median number of commits and distinct authors across the years in Blender are \MedianBlenderCommits\ and \MedianBlenderDistinctAuthors, respectively, while in \llvm, it is \MedianLLVMCommits\ and \MedianLLVMDistinctAuthors.

One more explanation is related to organizational self-correcting behavior.
Each product we study has an informal or formal core team.
The core team contains the most active or senior project members.
Based on our experience with \oss, the \cov\ falling into a specific range can also result from core team members ensuring that \cor s get a timely response and the number of pending issues decreases.

\section{Threats to validity}

Our study is subjected to a specific category of threats.
One threat relates to application of our findings in other contexts or \emph{external validity}~\cite{shull_guide_2008}.
The projects that we  investigate are all \oss.
The incentive structure in \oss\ development is different from industrial projects.
In addition, because of the \cor\ granularity that we target, the project selection is limited to the ones that use \phab\ for code collaboration.
We do not recommend generalizing these results without further replication in the target environment.

Another threat relates to \emph{internal validity}.
This threat type relates to  the interpretation of the results and if correct conclusions are drawn from the data.
We use the standard recommended nonparametric statistical apparatus to draw our conclusions.
The metrics such as rolling \num{30}-day means and median are widely used in quantitative finance as trend indicators~\cite{hyndman_2011}.
We corroborate our findings by calculating \num{90}-day and all-time medians that indicate a similar trend.

\section{Conclusions and future work}

We investigate the trend for \cov\ in four major \oss\ projects: Blender, \fbsd, \llvm, and \mo.
Our analysis is based on \InitialPhabricatorCount\ \cor s that span, on average, \AverageProjectAge\ years of development activity per project.
We find that \emph{\cov\ does not decrease over time}.
While the size of the code base in these projects increases on median between \AnnualSLOCIncrease\ annually, the \cov\ either stays the same or slightly improves.

We intend to replicate our findings for other code collaboration tools such as \ger\ and \gh.
While they do not provide the same granularity level as \phab, the findings will help invalidate or strengthen our claims.
In addition, our observation about the \cov\ falling into a specific range is worth additional research.

\section{Data Availability}

The \phab\ data, various {R} scripts that are used to perform statistical analysis, and relevant \textsc{SQL} queries are available on Figshare.\footnote{\protect\url{https://figshare.com/s/4558d92adc8d5d262bd6}}

\bibliographystyle{IEEEtran}
\bibliography{slowing-or-speeding}

\end{document}